\documentclass[12pt,preprint]{aastex}

\usepackage{graphicx,epsfig,natbib,color}
\usepackage{lscape}
\usepackage{graphicx}
\usepackage{epsfig}
\usepackage{natbib}

\begin{document}
\title{A Comparative Study of Confined and Eruptive Flares in NOAA AR 10720}

\author{X. Cheng\altaffilmark{1,2,3}, J. Zhang\altaffilmark{2}, M. D. Ding\altaffilmark{1,3}, Y. Guo\altaffilmark{1,3}, J. T. Su\altaffilmark{4}}

\affil{$^1$ Department of Astronomy, Nanjing University, Nanjing 210093, China} \email{dmd@nju.edu.cn}

\affil{$^2$ School of Physics, Astronomy and Computational Sciences, George Mason University, 4400 University Drive, MSN 6A2, Fairfax, VA 22030, USA}

\affil{$^3$ Key Laboratory for Modern Astronomy and Astrophysics (Nanjing University), Ministry of Education, Nanjing 210093, China}

\affil{$^4$ National Astronomical Observatories, Chinese Academy of Sciences, Beijing 100012, China}

\begin{abstract}
We investigate the distinct properties of two types of flares: eruptive flares associated with CMEs and confined flares without CMEs. Our sample of study includes nine M and X-class flares, all from the same active region (AR), six of which are confined and three others are eruptive. The confined flares tend to be more impulsive in the soft X-ray time profiles and show more slender shapes in the EIT 195 \AA\ images, while the eruptive ones are of long-duration events and show much more extended brightening regions. The location of the confined flares are closer to the center of the AR, while the eruptive flares are at the outskirts. This difference is quantified by the displacement parameter, the distance between the AR center and the flare location: the average displacement of the six confined flares is 16 Mm, while that of eruptive ones is as large as 39 Mm. Further, through nonlinear force-free field extrapolation, we find that the decay index of the transverse magnetic field in the low corona ($\sim$10 Mm) have a larger value for eruptive flares than that for confined one. In addition, the strength of the transverse magnetic field over the eruptive flare sites is weaker than that over the confined ones. These results demonstrate that the strength and the decay index of background magnetic field may determine whether or not a flare be eruptive or confined. The implication of these results on CME models is discussed in the context of torus instability of flux rope.

\end{abstract}

\section{Introduction}
Flares and coronal mass ejections (CMEs) are the two most energetic events that occur in the solar corona. They can together release a vast amount of mass, magnetic field and energetic particles into the outer space, which may severely disturb the space environment \citep{gosling93,webb94}. Albeit the importance of the topic and the efforts made in the past, the intrinsic physical relationship between flares and CMEs remains an elusive issue in solar physics \citep{kahler92,gosling93,schrijver09}. Recent studies have shown that flares are closely related with CMEs in their energetic processes when they are associated \citep{zhang01,zhang04,qiu04,temmer08}. \citet{zhang01,zhang04} found that flare-associated CMEs generally undergo three distinct phases of kinematic evolution: the initiation phase, impulsive acceleration phase, and propagation phase, which are closely coincided with the three phases of the associated flares: the pre-flare phase, flare rise phase, and flare decay phase in soft X-rays, respectively \citep[also see][]{bur04,vrsnak05,cheng10a}. \citet{qiu04} and \citet{temmer08,temmer10} studied the temporal relation between CME acceleration and the hard X-rays flux of the associated flares and found that they are also tightly correlated. \citet{cheng10b} made a statistical study of 247 CMEs associated with M- and X-class $GOES$ soft X-ray flares from 1996 to 2006, and found that the CMEs associated with flares with long-decay time are more likely to have a positive post-impulsive-phase acceleration. It is generally believed that, when associated, flares and CMEs may be different manifestation of the same magnetic energy releasing process in the corona, possibly through magnetic reconnection \citep{lin00,priest02,zhang06,mari07,temmer08,cheng10a}.

On the other hand, it has been shown that not all flares are associated with CMEs \citep{and03,ya05}. \citet{ya05} found the possibility of the flares being associated with CMEs increases with the X-ray flare magnitude. In order to distinguish the flares without accompanying CMEs from the ones with CMEs, \citet{svestka92} named them as ``confined flares" and ``eruptive flares", respectively. \citet{wang07} compared the magnetic properties between four confined and four eruptive X-calss flares from a variety of active regions (ARs). They found that the confined ones usually occur close to the magnetic center of the AR; while the eruptive ones generally occur far from the magnetic center, e.g., at the edge of the AR. Using potential field source surface model, they also calculated the ratio of the magnetic flux in the lower corona to that in the higher corona and found that the ratio is lower for the confined flares than the eruptive ones. \citet{torok05}, \citet{kliem06}, \citet{fan07}, and \citet{olmedo10} investigated the torus instability of flux rope structures. It was found that the decay of the background magnetic field with height is a critical factor in determing whether the instability of the flux rope can result in an eruption or not, i.e., the decay index should be larger than a critical value in order to have an successful eruption. \citet{liu08} studied ten events from different ARs, consisting of four failed eruptions, four eruptions due to kink instability, and two eruptions due to torus instability. They calculated the decay index of the background transverse magnetic field in the source ARs and found that the decay index for successful eruptions is larger than that for failed eruptions.

The above studies indicate the importance of background magnetic fields in determining the eruption of a flare, in particular, the decay index of the magnetic field. In this paper, we further investigate this issue using two types of flares originating from the same AR. This is a unique approach since it removes other factors contributing to the eruption, such as the complexity and scale size of the region. It should also be noted that, in previous studies, the decay properties of the background magnetic field of the ARs were investigated using the potential field models. Nevertheless, many studies revealed that electrical currents are ubiquitous in flare productive ARs \citep[e.g.,][]{schrijver08,schrijver09}. The currents make the coronal magnetic field non-potential persistently, especially in the low corona close to the magnetic polarity inversion line (PIL) of the region. \citet{thalmann08} and \citet{cheng10c} found that the magnetic field configuration of the ARs, even after a major eruption, is not necessarily potential. \citet{schrijver09} showed that using the potential field method to calculate the background magnetic field is an oversimplification. Therefore, it is useful to investigate the properties of the background magnetic field using non-linear force free field (NLFFF) models. In this paper, we make such study of six confined and three eruptive flares that all from the NOAA AR 10720. In Section 2, we present the observations and data analysis. The distinct properties of the confined and eruptive flares are shown in Sections 3 and 4. We make discussions and summary in Section 5.

\section{Observations and Data Analysis}

\begin{figure*} 
\centerline{\hspace*{0.0\textwidth}
               \includegraphics[width=0.8\textwidth,clip=]{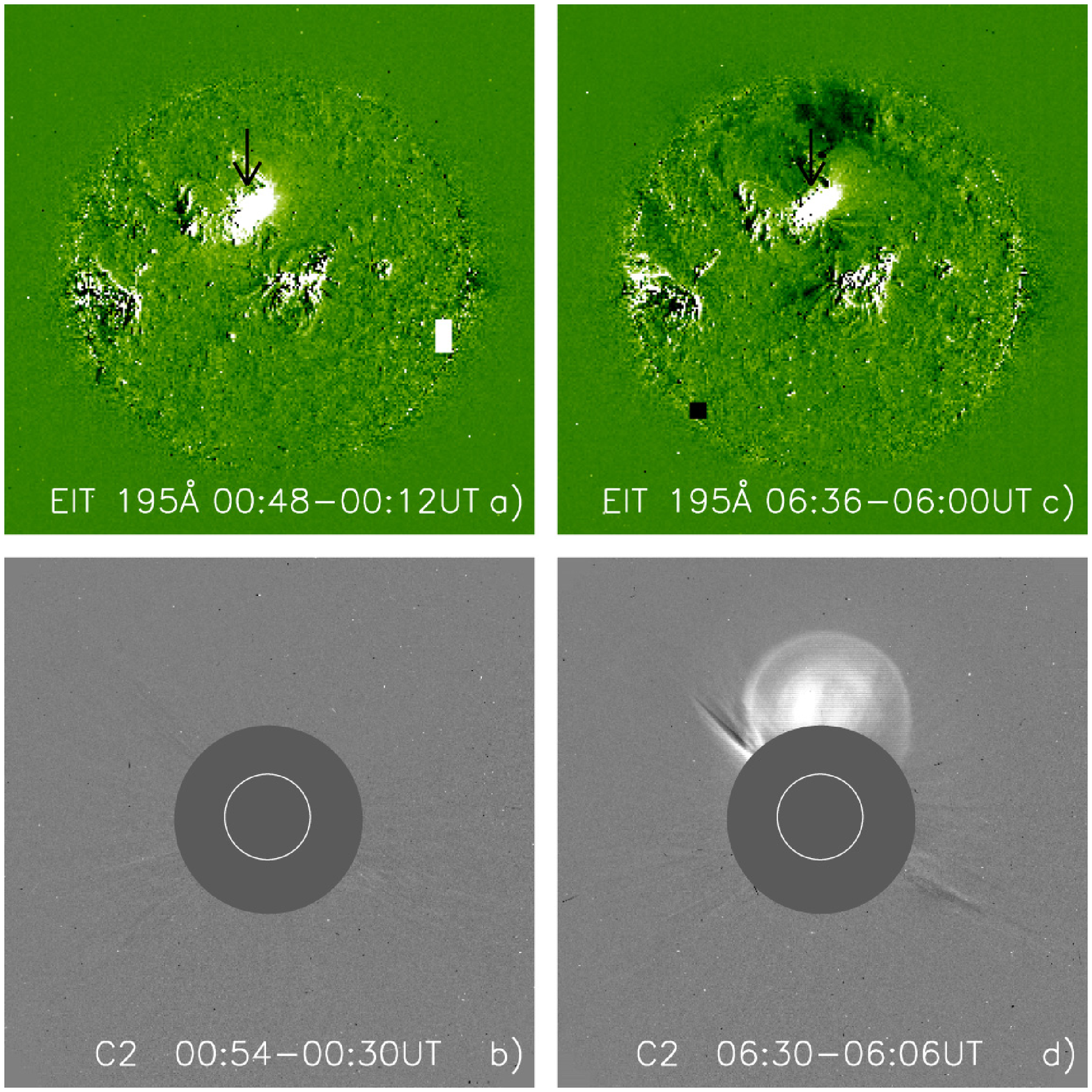}
            }
\vspace{-0.0\textwidth}    
\caption{Running difference images of EIT 195\AA\ and LASCO C2 for a
confined flare on 15 January (left) and a eruptive flare on 15
January (right). Two flares location are shown by the arrows.}
   \label{cme}
   \end{figure*}

NOAA AR 10720 was a flare-productive AR that appeared first in the eastern limb on 2005 January 10 and finally disappeared in the western limb on January 23. During its 14-day-long trespassing across the front-disk of the sun, AR 10720 produced 17 M-class and 5 X-class soft X-ray flares
based on the record of $GOES$ satellites that provide the full disk soft X-ray emission in 1--8 \AA. In order to determine the confinement or eruptiveness of these flares, we visually inspect the images obtained by the Extreme-ultraviolet Imaging Telescope (EIT; \citet{Del95}) and the Large Angle Spectroscopic Coronagraph (LASCO; \citet{bru95}), both of which are on board the \textit{Solar and Heliospheric Observatory} (\textit{SOHO}) spacecraft. A flare is taken as an eruptive one if and only if the transient flare brightening seen on the EIT is associated with an apparent CME on the LASCO. The flare peak time should be within $\pm$60 minutes of the first appearance time of CME on C2 coronagraph and the flare coordinate must lie within $\pm$60$^{\circ}$ of the position angle of CME. An eruptive flare is also very often associated with a large scale dimming surrounding the flare region seen on the EIT. On the other hand, a confined flare is not associated with any CME. Such a visual-inspection method using both EIT and LASCO is regarded as the most reliable way that checks whether or not a flare is associated with a CME \citep[also see,][]{wang07,cheng10b}. As an example, we show the EIT running-difference images of a confined flare along with an eruptive flare in Figure \ref{cme}, as well as the complementing LASCO images. The confined flare occurred at 00:48 UT on 15 January (Figure \ref{cme}a) but without CME features shown in the C2 field of view (FOV) (Figure \ref{cme}b); while, the eruptive flare occurred at 06:36 UT on the same day was associated with a large-scale dimming surround the location (Figure \ref{cme}c) and an obvious CME in C2 (Figure \ref{cme}d).

\begin{figure*} 
\centerline{\hspace*{0.0\textwidth}
               \includegraphics[width=0.33\textwidth,clip=1]{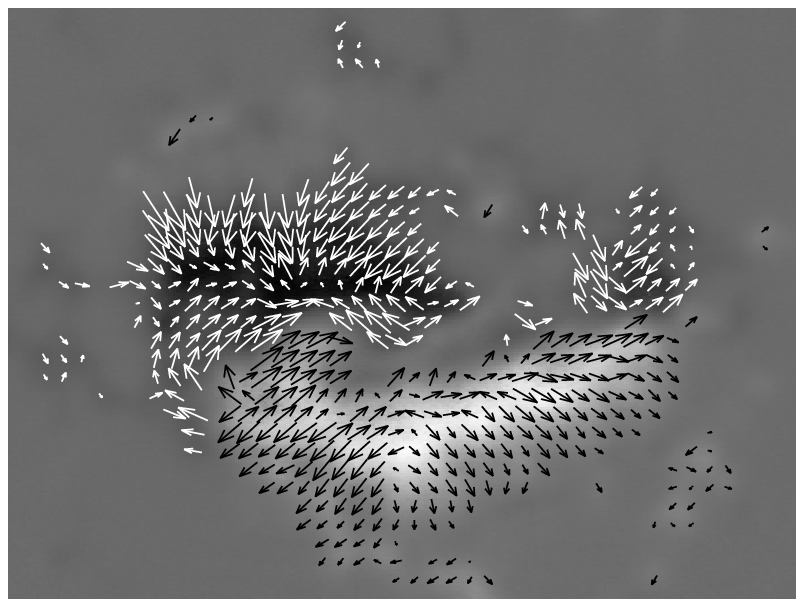}
               \includegraphics[width=0.33\textwidth,clip=]{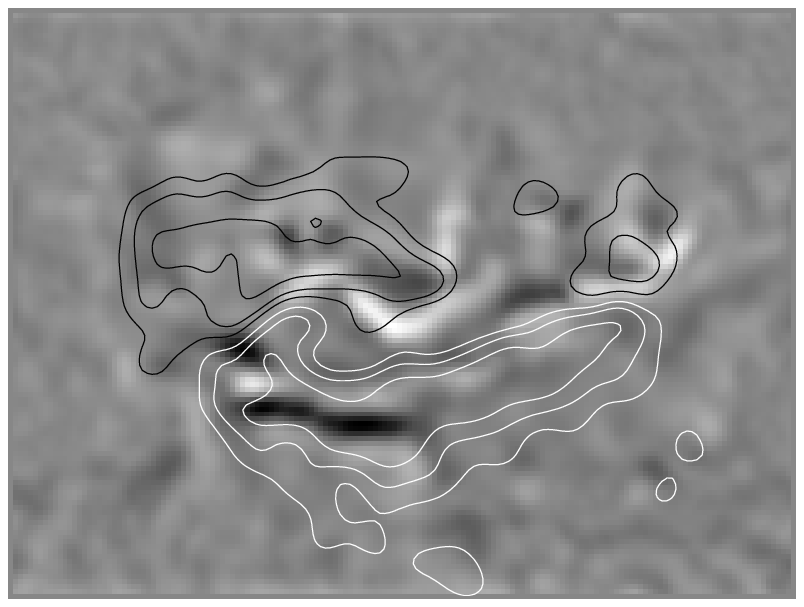}
               \includegraphics[width=0.33\textwidth,clip=]{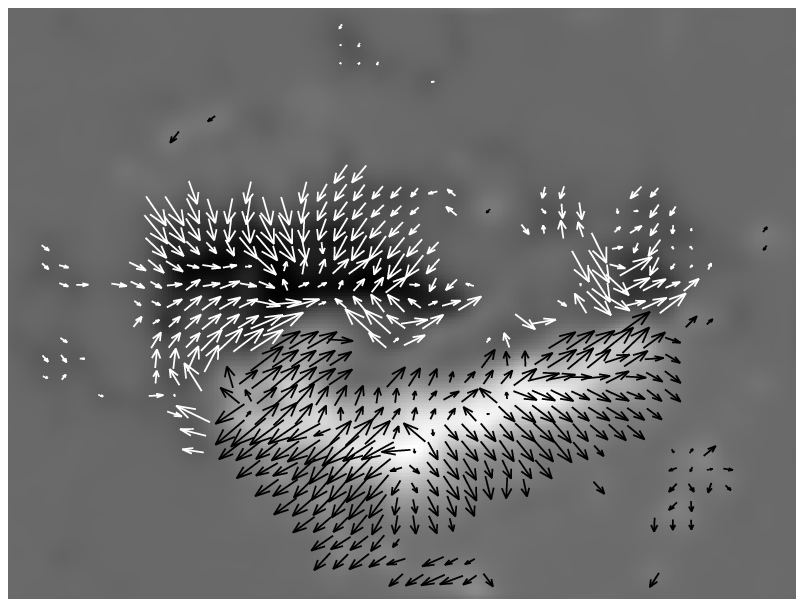}
            }

     \vspace{-0.0\textwidth}    
\caption{Left: vector magnetograms at 04:39 UT on 15 January taken
by the SFMT at HSOS with the 180$^{\circ}$ ambiguity removed.
Middle: map of the vertical current density $j_{z}$ overlayed by the
contours of the line-of-sight field, the color scale is from black
(negative) to white (positive). Right: preprocessed vector
magnetogram used for NLFFF extrapolation.}
   \label{vector}
   \end{figure*}

Magnetic field data of this AR are provided by the Solar Magnetic Field Telescope (SMFT) at the Huairou Solar Observing Station (HSOS) in China. It made many instances of measurements of the vector magnetic field of NOAA AR 10720. In order to minimize the projection effect, we select only the magnetograms in which the AR was close to the disk center, i.e., within 30$^{\circ}$ of the central meridian. Within this longitudinal band, we obtain six confined (F1--F6) and three eruptive flares (F7--F9), which constitute our sample of study in this paper, as shown in Table \ref{tb1} (details will be discussed in Section 3). The spatial resolution of the magnetograms from SMFT is 0.7$''$; the FOV is 225$''\times$ 168$''$ and thus well covers the area of the AR. The sensitivity of the line-of-sight field and the transverse field are better than 20 and 150 G, respectively \citep{li02}. The 180$^{\circ}$ azimuthal ambiguity of the transverse field is resolved through comparing the angles between the observed transverse field and the extrapolated linear force-free field (LFFF) \citep{wang97,wang01,metcalf06}. The extrapolation of LFFF is based on the observed line-of-sight field as the bottom boundary. The left panel of Figure \ref{vector} shows the vector magnetogram of the AR, whose 180$^{\circ}$ ambiguity has been removed. The AR first appeared as a simple dipole. As the successive emergence of new fluxes in the AR, the magnetic shear was obviously enhanced, which resulted in complicated magnetic field structure \citep{cheng10c}. Nevertheless, the PIL was generally a single elongated line. The middle panel shows the distribution of the vertical current density on the photosphere, and right panel shows the pre-processed magnetogram necessary for the NLFFF calculation as discussed in the next paragraph.

Using the observed vector magnetogram as the bottom boundary, we extrapolate the three-dimensional (3D) coronal magnetic field through the optimization-based NLFFF model. The model was first proposed by \citet{wheatland00} and implemented by \citet{Wie04}. Prior to the extrapolation, the bottom boundary data have to be treated by a preprocessing procedure proposed by \citet{Wie06} in order to reduce the inconsistency between the forced photospheric magnetic field and the force-free assumption in the models, as well as reducing the noise of the observed magnetic field. This preprocessing of the input magnetogram minimizes the net force and torque of the photospheric magnetic field. It also maintains the consistency between the final preprocessed data and the measured data if the flux balance condition in the FOV is met. However, limited by observations, the
flux balance condition is usually not fully satisfied. \citet{derosa09} did the experiment of embedding vector magnetograms into the line-of-sight magnetograms of a larger FOV, in an effort to achieve the flux balance condition. Nevertheless, they showed that the unknown transverse magnetic fields in the larger FOV often cause an inconsistent boundary condition. \citet{guo10} further argued that applying the preprocessing procedure on the magnetic field in the original FOV is probably a better approach than embedding it in a larger FOV where the transverse field information is missing. We have checked the flux balance condition $\epsilon$ of NOAA AR 10720 and found that it is satisfied to a good degree (Table \ref{tb2}).

\begin{figure*} 

\centerline{\hspace*{0.\textwidth}
               \includegraphics[width=1.0\textwidth,clip=]{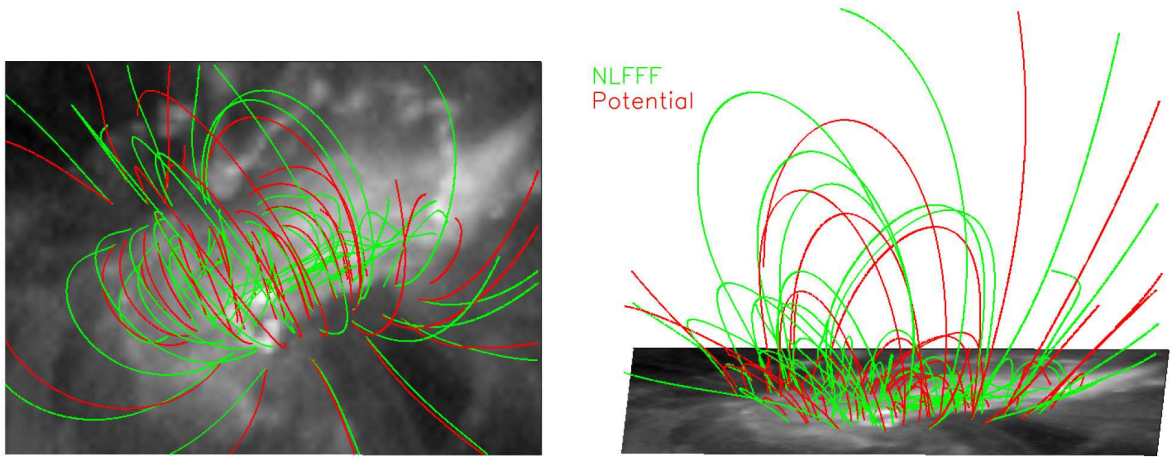}
            }
     \vspace{-0.\textwidth}    
\caption{Extrapolated 3D NLFFF (green) and potential field (red) configuration at 06:41 UT on 15 January. Left: top
view; right: side view. The background is EIT image at 06:00 UT.}
   \label{3dfield}
   \end{figure*}

Further, through the extrapolated 3D magnetic field, we inspect the isolation condition, which is also required for the 3D field extrapolation \citep{aly89}. We calculate the two isolation criteria, $\epsilon_{in}$ and $\epsilon_{out}$ (Table \ref{tb2}), defined as the ratio of the passing magnetic flux from the four sides and top boundaries to that from the bottom boundary for inward and outward magnetic fields, respectively. Zero values of $\epsilon_{in}$ and $\epsilon_{out}$ would indicate that the extrapolated region is perfectly magnetically isolated. However, we find that the calculated values deviate from zero, indicating one source of uncertainty of the calculated coronal magnetic field. Nevertheless, we believe that the NLFFF model is a better approach than the PFSS model for studying the observations in this paper. The resulted coronal magnetic field lines from the NLFFF and potential field models are shown in Figure \ref{3dfield}, from which it is clear that the NLFFF model reproduces the strong sheared core field in the low corona underneath the arched overlying field (green lines). The strong sheared and elongated field lines along the PIL indicate there might exist helical flux rope structure. On the other hand, one can see from red lines in Figure \ref{3dfield}, the potential model can only reproduce the arched overlying field but not helical magnetic structure. The further properties of the overlying field will be discussed in section 4.

\section{Distinct-Properties of Confined and Eruptive Flares}
The overall properties of the confined and eruptive flares in our sample are summarized in Table 1, including the occurring date, onset time, rise time, duration, location in heliographic coordinates, and $GOES$ X-ray class, as well as the associated CMEs' velocity and angular width.  It is obvious that the confined flares (F1--F6) tend to be more impulsive, or of short duration. Their rise time and duration have an average of 11 and 23 minutes, with a maximum of 21 and 40 minutes, respectively. Whereas, for the eruptive flares (F7--F9), they are all long-duration events (LDEs); the rise time and duration have an average of 84 and 112 minutes, with a minimum of 37 and 66 minutes, respectively. This difference in the flare duration is remarkable, considering that the flares all originated from the same AR, thus shared the same global magnetic field environment. The cause of the difference must be of the physical process in the local scale. It has been believed that the long-duration flares may result from the continual or extended magnetic reconnection driven by a positive feedback between the CME eruption and a fast magnetic reconnection inflow \citep{zhang06,cheng10a}; such a positive feedback does not exist in confined flares. The speed of these LDE-associated CMEs all exceed 2000 km s$^{-1}$, which may be the result of the combination of a strong acceleration rate and the long duration of such strong acceleration. This kind of extremely fast CMEs are rather rare in general CME population \citep[e.g.,][]{wang08}. Recently, \citet{cheng10c} studied in detail the F7 and F8 events and found that F8 was caused by the re-flaring of the post-flare loop system of F7; the flares are believed to be driven by the continuous emergence of magnetic flux. Also note that, F9 event has an extremely long duration of 188 minutes. In fact, it was associated with two fast halo CMEs, which erupted consecutively within a short time.

\begin{figure*} 
     \vspace{-0.0\textwidth}    
     \centerline{\hspace*{0.0\textwidth}
               \includegraphics[width=0.95\textwidth,clip=]{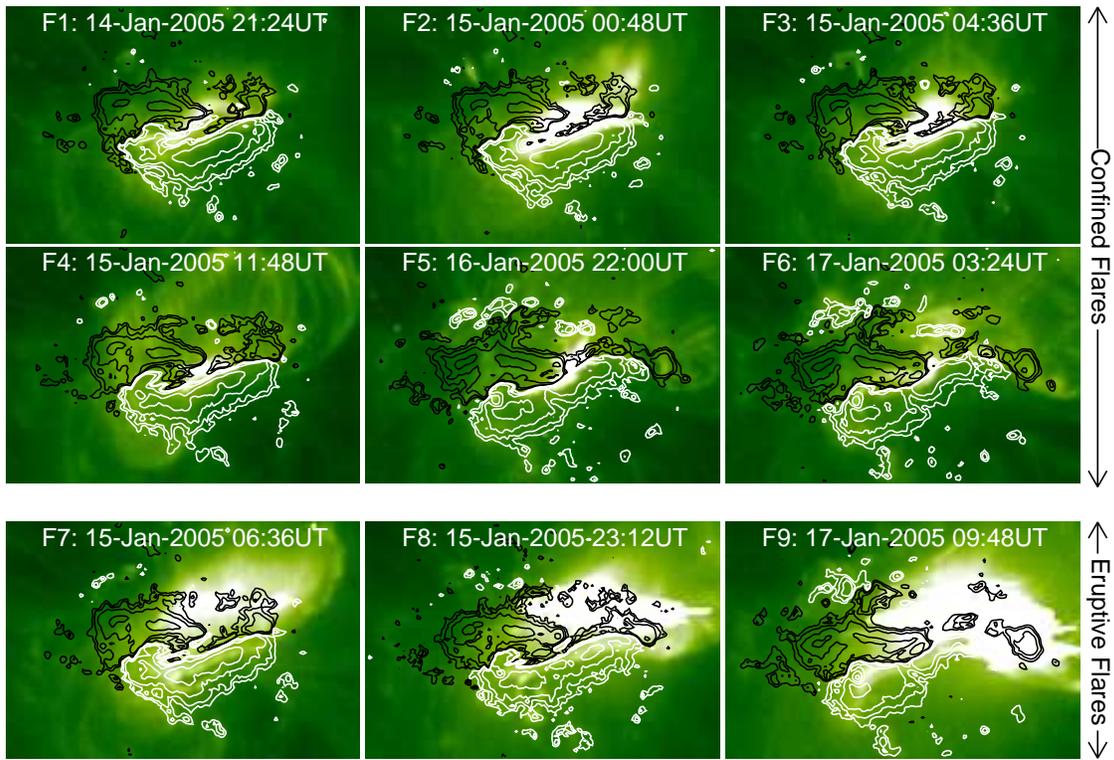}
              }
     \vspace{0.0\textwidth}   

\caption{EIT 195 \AA\ images superimposed with contours of the
line-of-sight magnetograms from MDI. The top two rows are for
confined flares and the bottom row is for eruptive flares.}
   \label{eit}
   \end{figure*}

As seen in the EIT images, the confined and eruptive flares show quite different morphology (Figure \ref{eit}). All confined flares show a more slender and compact brightening region along the PIL, whereas eruptive flares show broader and more extended brightenings.

\begin{figure*}  

\centerline{\hspace*{0.0\textwidth}
               \includegraphics[width=0.8\textwidth,clip=]{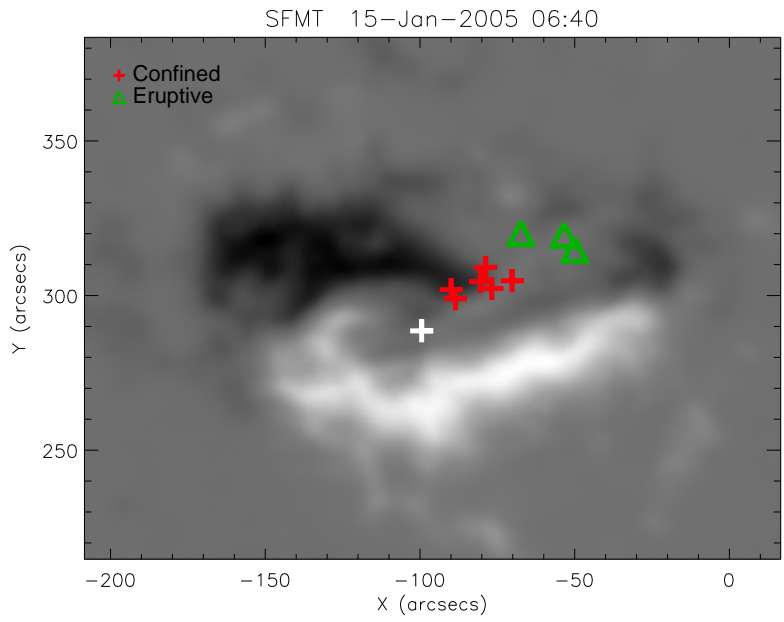}
            }

     \vspace{-0.0\textwidth}    
\caption{Line-of-sight magnetograms with a FOV of
225$''\times$ 168$''$ taken by the SMFT at HSOS. The white plus
indicates the magnetic flux-weighted center, the red and green pluses denote
the intensity-weighted centers of the confined and eruptive flares, respectively.}
   \label{distance}
   \end{figure*}

More interestingly, the two kinds of flares clustered at different locations with respect to the center of the AR (Figure \ref{distance} and Table \ref{tb1}), though they all located along the same general PIL of the AR. A parameter $D$ is defined as the distance between the intensity-weighted flare brightening center and the magnetic flux-weighted AR geometric center, which can well quantify the relative position of the flare \citep[also see,][]{wang07}. We find that the six confined flares tend to occur near the magnetic flux-weighted center with an average $D$ of 16 Mm and a maximum of 29 Mm. On the other hand, the three eruptive flares all occurred far from the AR center with an average $D$ of 39 Mm and a minimum of 31 Mm. Note that, the measurement error $\delta D$ of the parameter $D$ mainly comes from the uncertainty of the weak magnetic field measurement. Based on the uncertainty of $\sim$2\% of total magnetic flux of the AR and the AR characteristic scale length of 50 Mm, we estimate that $\delta D$ is about 1 Mm. Further, the spatial resolution of the magnetograms used in our study is $\sim$1 Mm (1.4$''$). Therefore the final $\delta D$ is estimated to be $\sim$2 Mm.

\begin{figure*}  
     \vspace{-0.0\textwidth}    
     \centerline{\hspace*{0.0\textwidth}
               \includegraphics[width=0.6\textwidth,clip=]{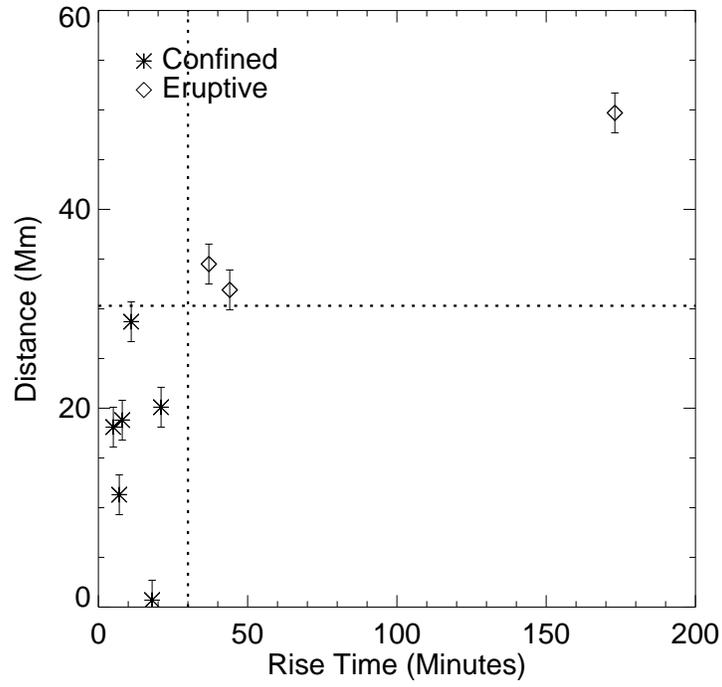}
              }
     \vspace{0.0\textwidth}   
\caption{Scattering plot of the distances \textit{D} between the intensity-weighted
flare center and the magnetic flux-weighted center versus the flare rise time for the two types of flares.}
   \label{scatter}
   \end{figure*}

To summarize the results presented above, we make a scattering plot of the distance versus the rise time for the two kinds of flares (Figure~\ref{scatter}). The two kinds of flares are separated into two groups by the dotted separating lines. Confined flares are cluttered in the lower-left quadrant with short distances and short rise times, while eruptive flares are located at the upper-right quadrant because of their large distances and long rise times. On the other hand, there is no flare in the other two quadrants. The results indicate that, while a flare could occur anywhere along the PIL, a CME-associated flare tend to occur at the outskirts of the source AR, and the confined flares close to the core of the AR. We will explain this phenomenon in the context of the magnetic property of the source region in the next section.

\section{Coronal Magnetic Properties of Confined and Eruptive Flares}
\begin{figure*}  
     \vspace{-0.0\textwidth}    
     \centerline{\hspace*{0.00\textwidth}
               \includegraphics[width=0.33\textwidth,clip=]{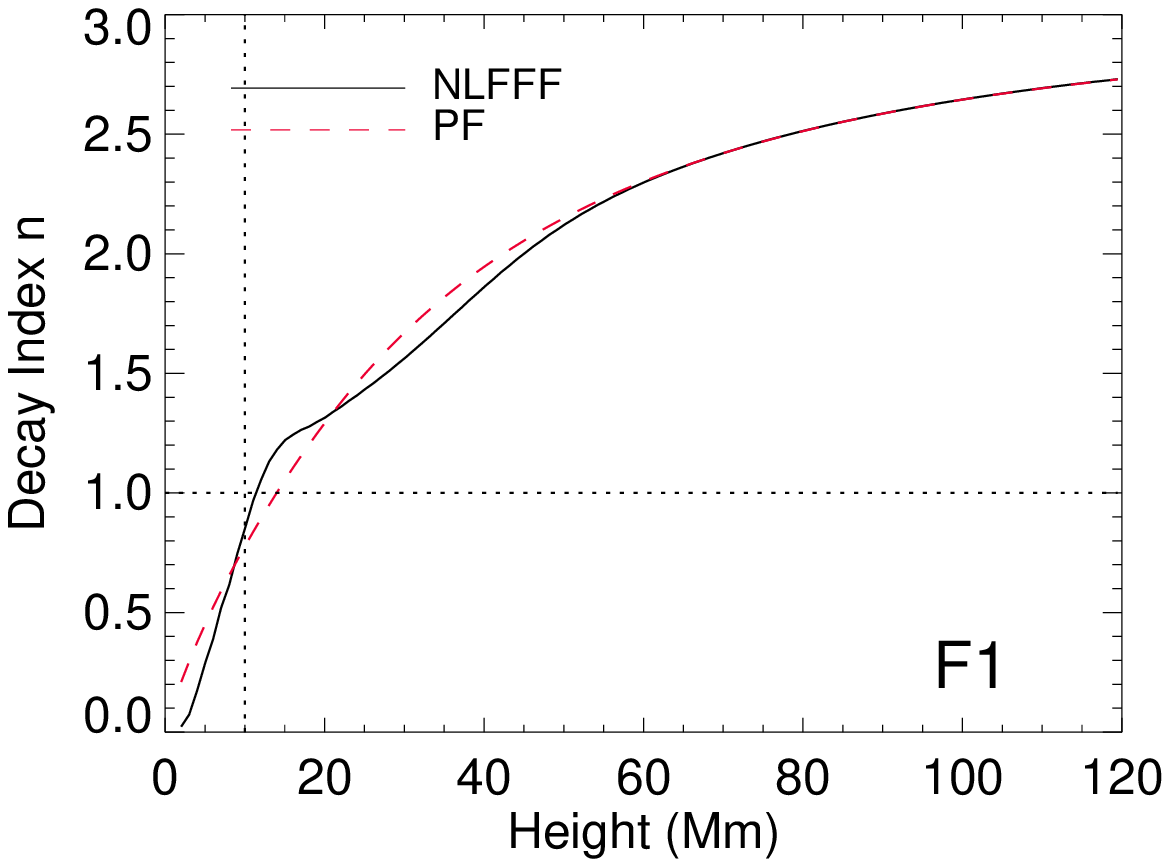}
               \includegraphics[width=0.33\textwidth,clip=]{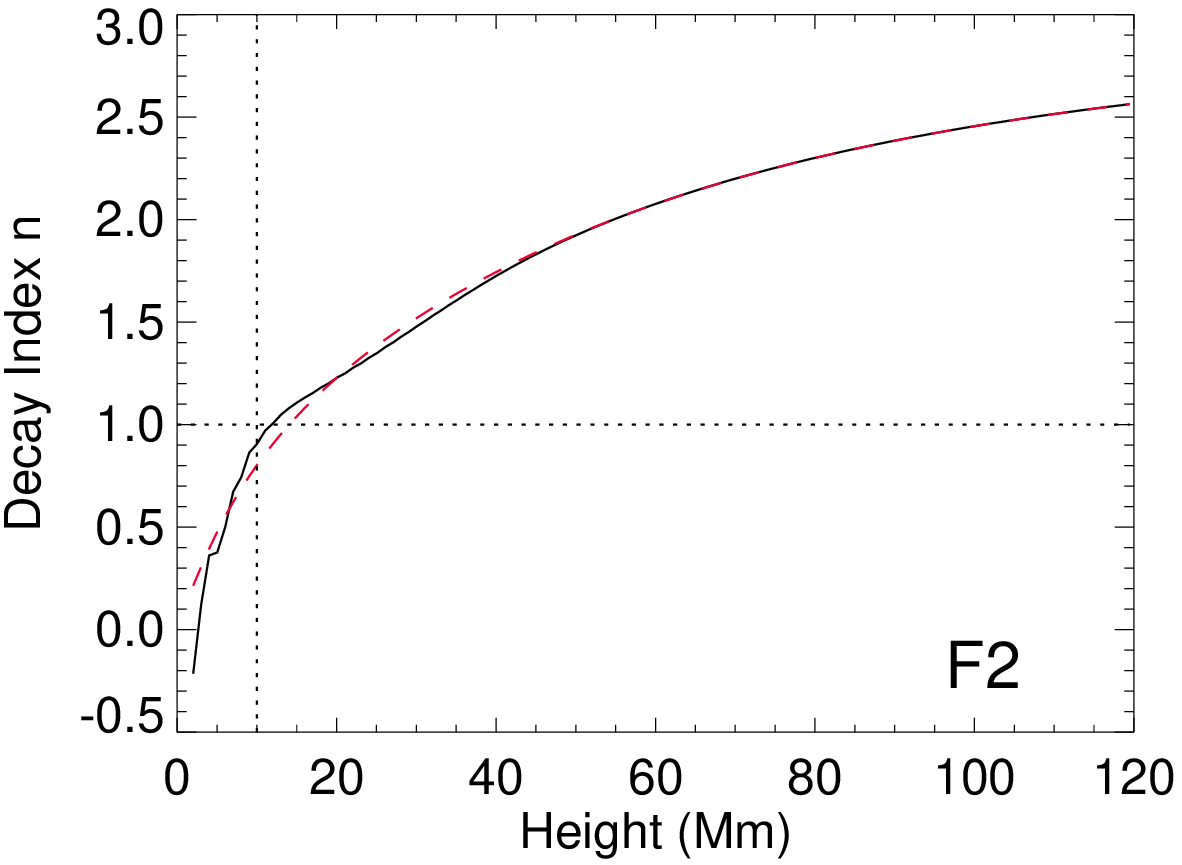}
               \includegraphics[width=0.33\textwidth,clip=]{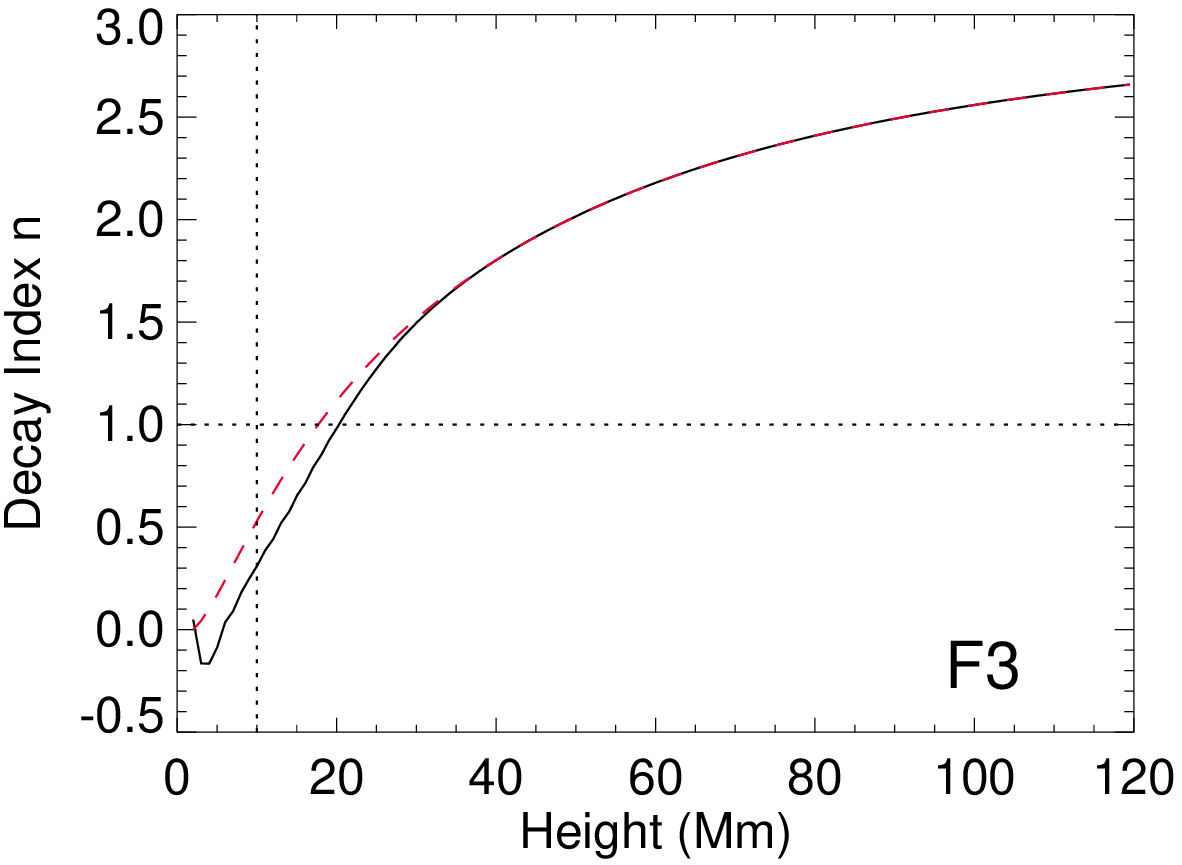}
               }
                    \centerline{\hspace*{0.00\textwidth}
               \includegraphics[width=0.33\textwidth,clip=]{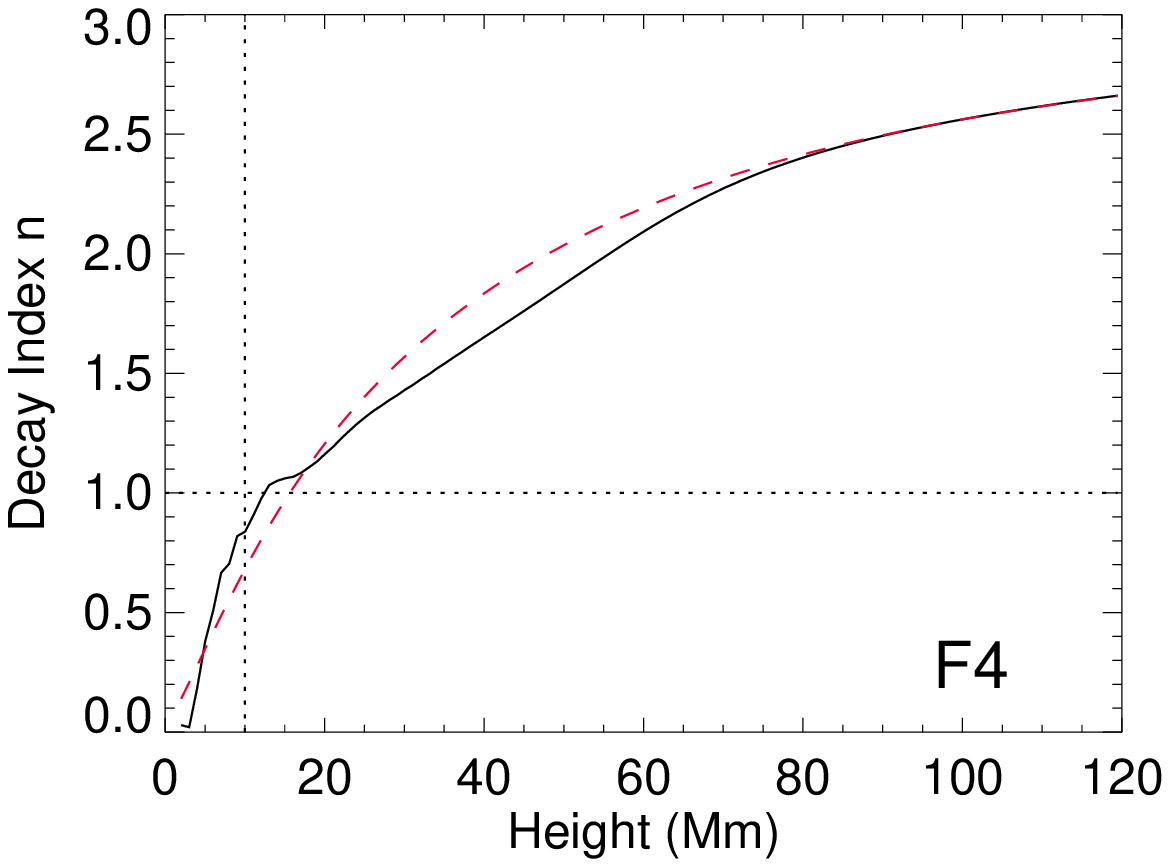}
               \includegraphics[width=0.33\textwidth,clip=]{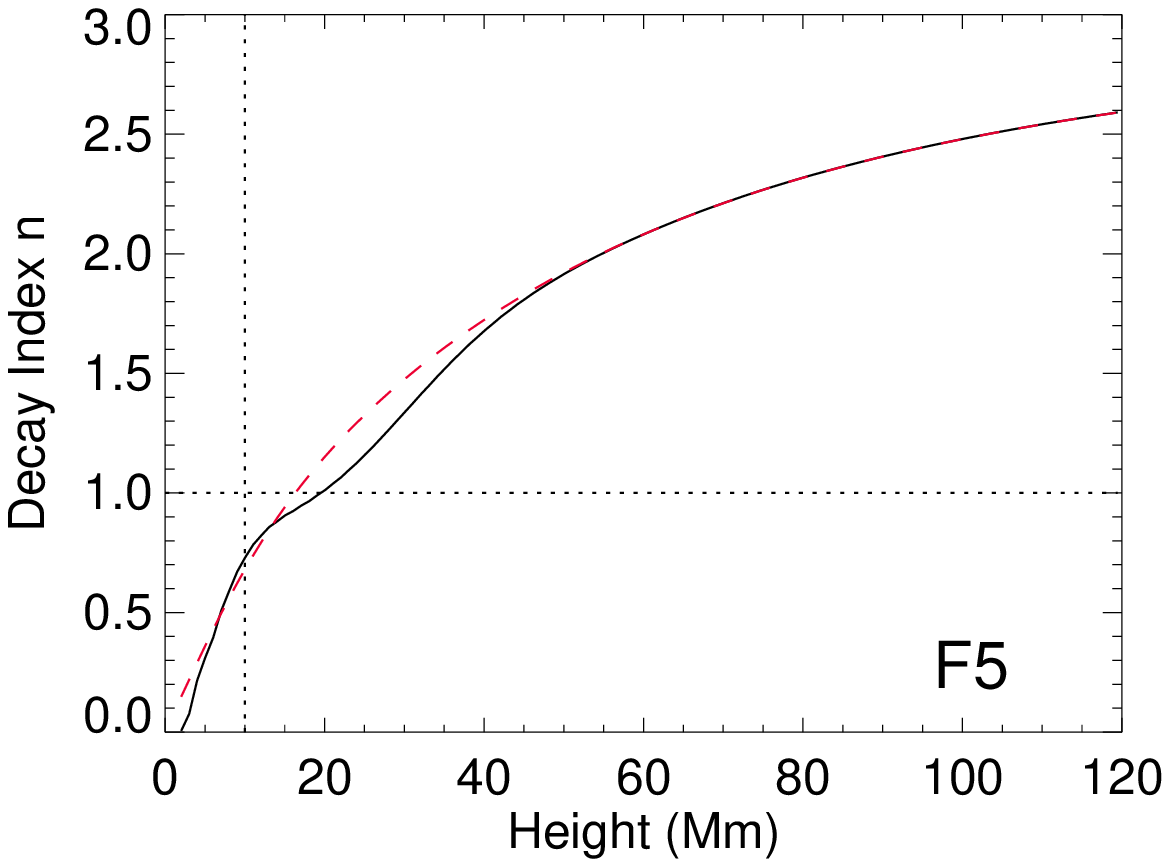}
               \includegraphics[width=0.33\textwidth,clip=]{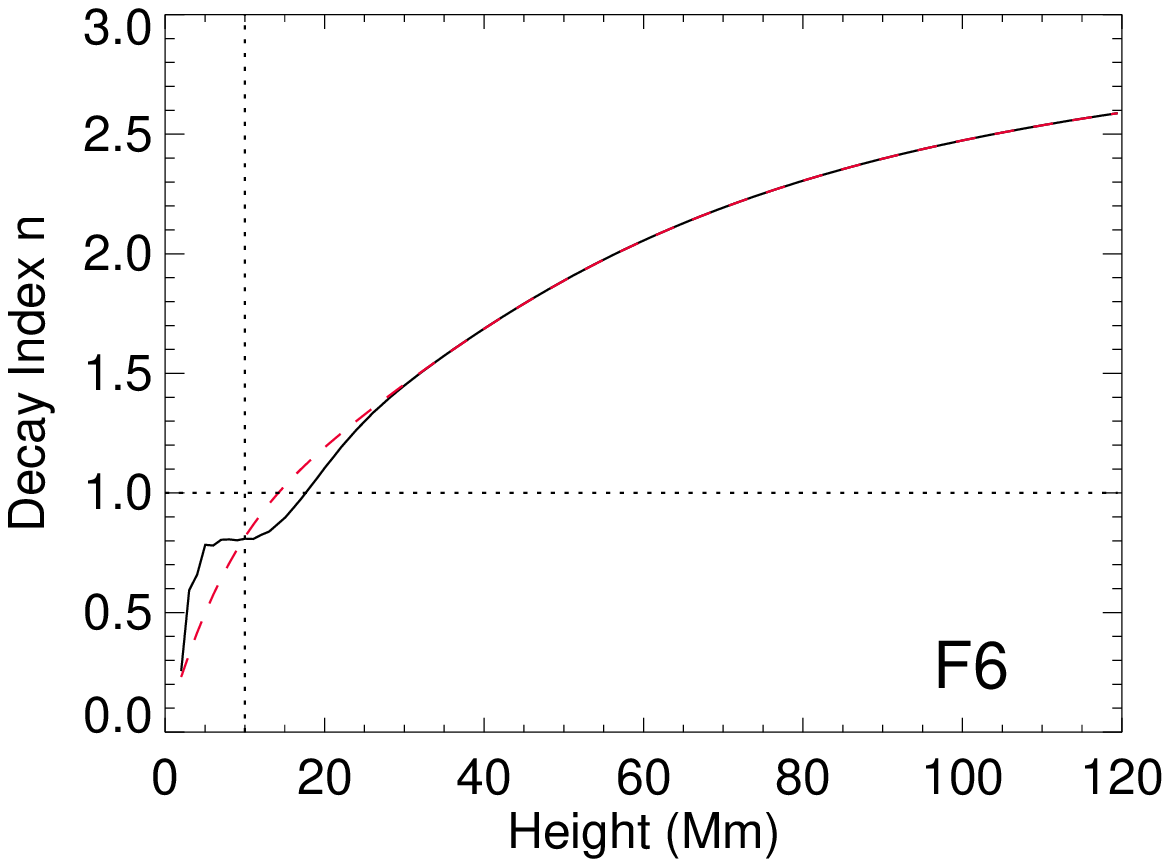}
               }
                    \centerline{\hspace*{0.00\textwidth}
               \includegraphics[width=0.33\textwidth,clip=]{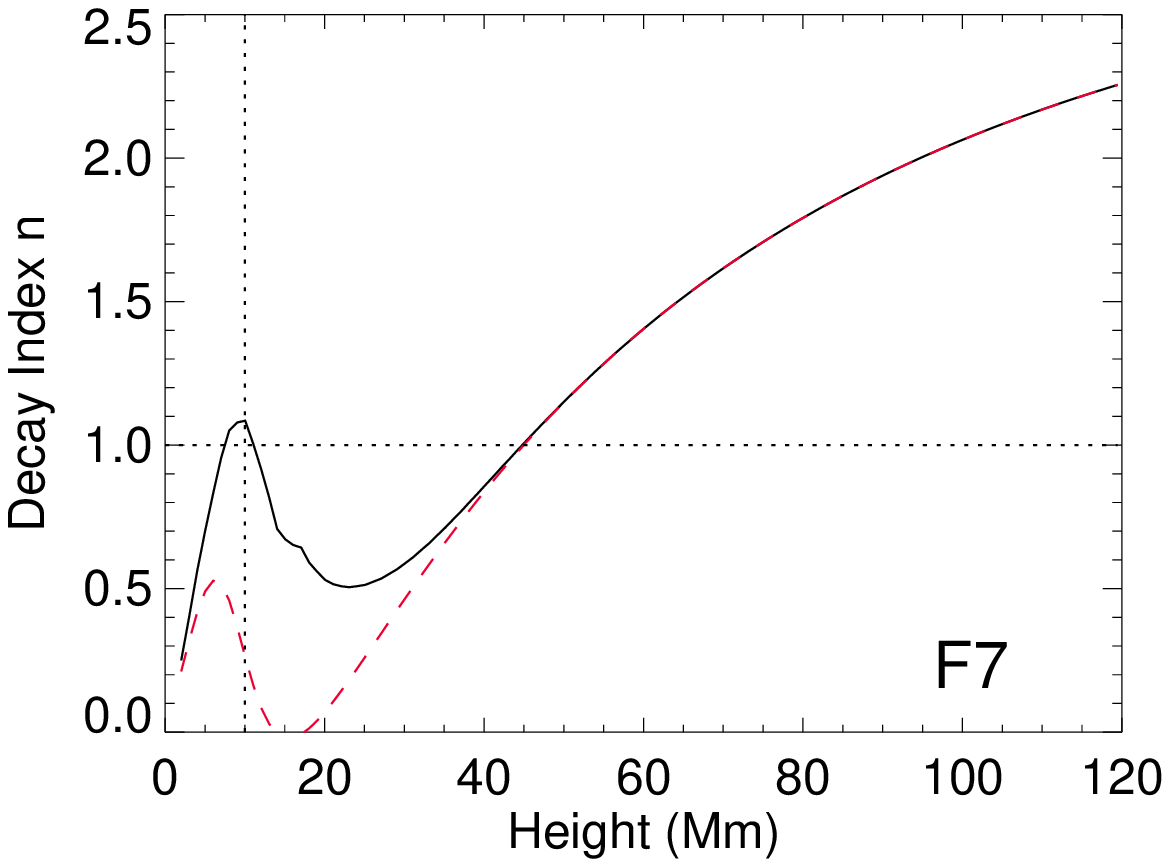}
               \includegraphics[width=0.33\textwidth,clip=]{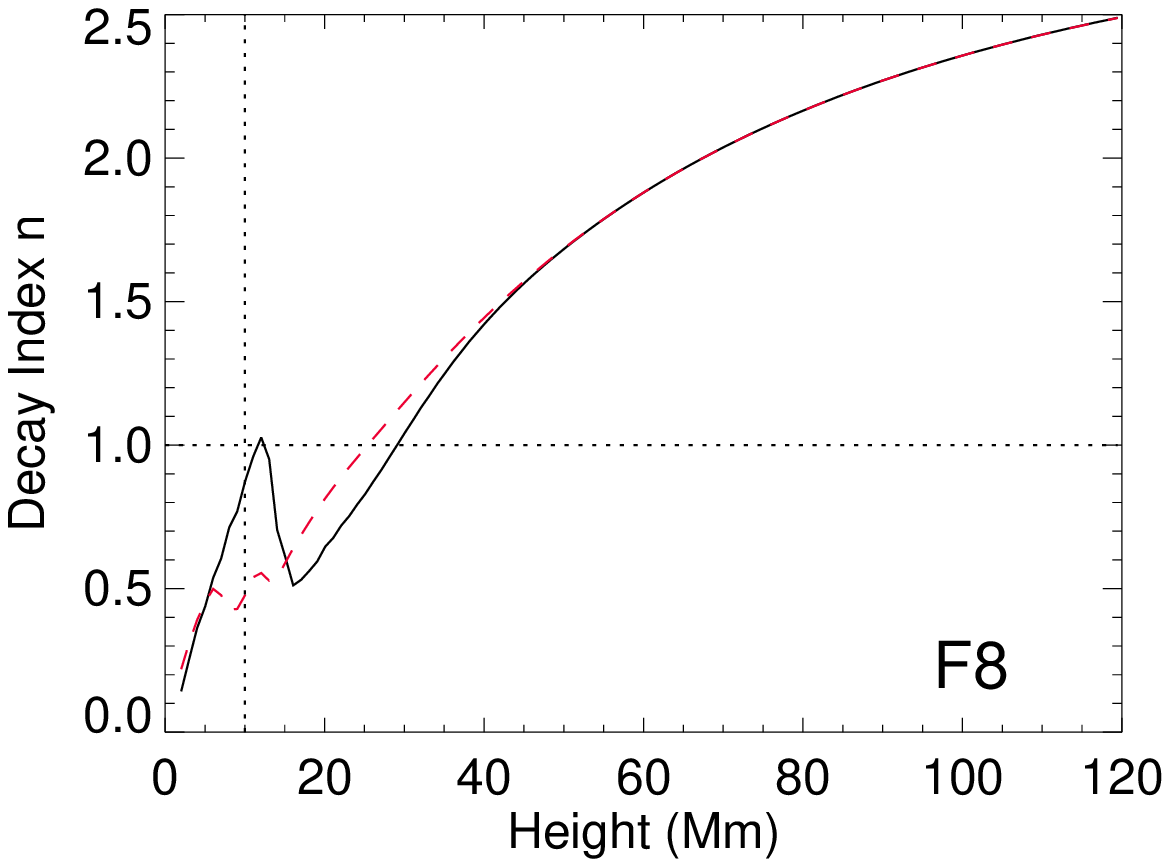}
               \includegraphics[width=0.33\textwidth,clip=]{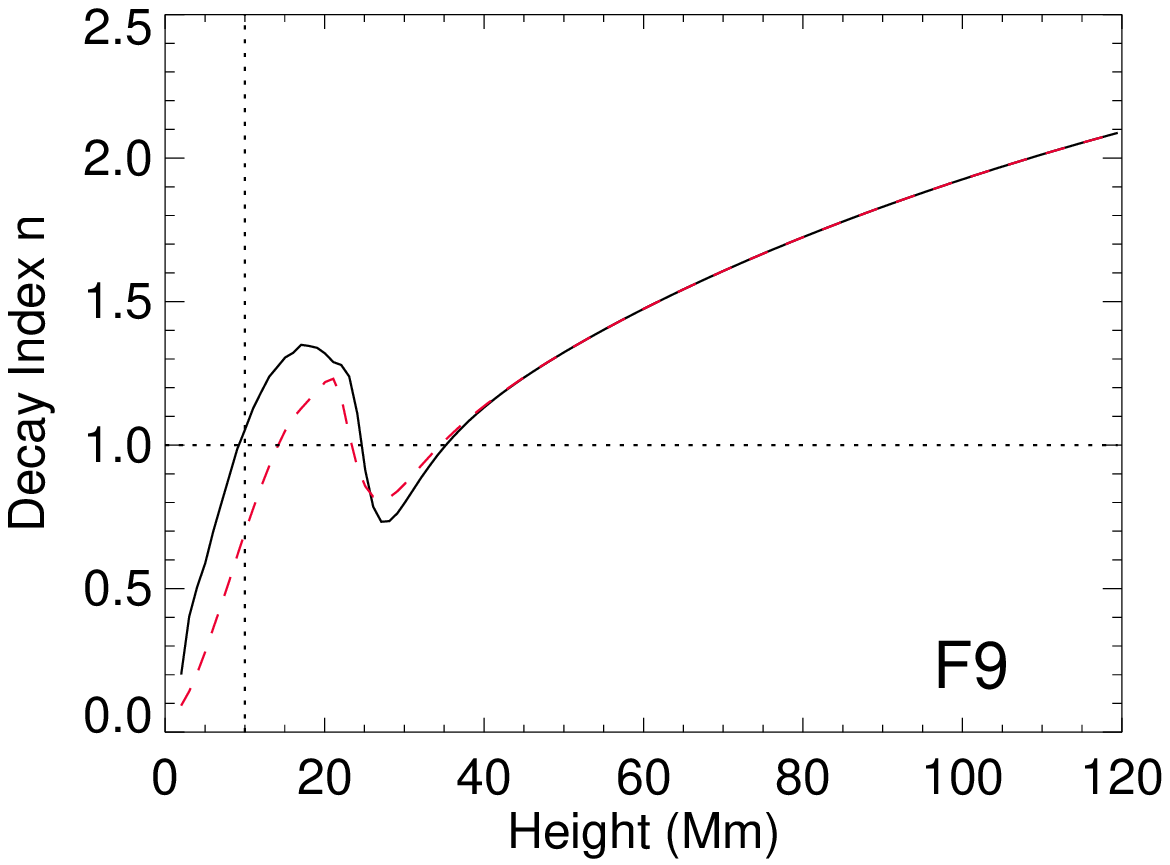}
               }

     \vspace{0.0\textwidth}   
\caption{Distributions of the transverse magnetic field decay index with the height for the nine flares studied. The back solid and red dotted lines
are from the NLFFF and the potential field model calculation, respectively. The vertical dotted line indicates the height of 10 Mm.}
   \label{index}
   \end{figure*}

The structure of coronal magnetic field is generally believed to play the key role in producing flares and/or CMEs. In particular, the background magnetic field gradient, the so-called decay index, has been regarded as an important parameter controlling the instability of the flux rope \cite[e.g.,][]{torok05,fan07,isenberg07,liu08,Aulanier10,olmedo10,torok10}.
In these models, the flux rope is the presumed magnetic structure prior to the CME eruption, which has a tendency to erupt due to the Lorentz self-force. However, the overlying background magnetic field provides the downward Lorentz force to keep the balance of the flux rope. If the decay index of the overlying magnetic field reaches a critical value, it results in torus instability \citep{torok05} or partial torus instability \citep{olmedo10} that leads to CME. The decay index is defined as
\begin{equation}
n=-\frac{\log B}{\log h},\\
\end{equation}
where \textit{$B$} denotes the background magnetic field strength at the geometrical height \textit{h} above the surface at the eruption region. While the flux-rope-based CME model is an ideal model that has not been commonly accepted by the solar community, we are inspired by the concept and believe that, in any case, it is worthy to explore the overlying magnetic field distribution, in particular, the coronal decay index, in order to understand the cause of two types of flares. For each event in our sample, we choose the vector magnetic field measurement prior to the flare time as the boundary condition to extrapolate the 3D coronal field, then calculate the decay index $n$ using equation (1). However, the measurement nearest the flare time is replaced if there was not observation available before the flare time. Figure \ref{index} shows the plots of the decay index \textit{n} varying with the height for the nine flare events studied. One can find that the decay index generally increases from 0 at the surface to 2.5 at the height of $\sim$80 Mm, the upper boundary of the calculation. The magnetic field used in the calculation is the transverse component \textit{$B_t$} of the vector magnetic field \textit{$B$} in the corona, since the line-of-sight component does not contribute to the downward constraining force. At each height, the decay index is an average value over the pixels of core flare brightening region, which is chosen as the intensity more than 80\% of the flare maximum value in corresponding EIT images. The black solid lines in Figure \ref{index} are calculated from the NLFFF model based on surface vector magnetic fields, while the red dashed lines are from the potential field model based on the surface line-of-sight magnetic fields. At the nominal critical decay index of the torus instability, i.e., $n_c$=1.5 \citep{torok05}, the height of the extrapolated magnetic field is $\sim$30 Mm at the center of the AR, while the height is $\sim$50 Mm at the outskirt of the AR. These numbers may put a theoretical upper limit for the height of the flux rope prior to the eruption: not higher than 30 Mm inside the AR, and not higher than 50 Mm at the outskirts. In general, at the same height of the corona, the decay index is higher in the center of the AR than that in the outskirts of the AR. Note that, in the torus instability models, the decay index is calculated from the background magnetic field, which is separated from the magnetic field induced by the current inside the flux rope. However, the extrapolated magnetic field in our observation-based calculation is the total magnetic field, making no difference between the background field constraining the flux rope and the induced field created by flux rope. It is difficult to separate the background magnetic field from the total magnetic field in the NLFFF model calculation (T{\"o}r{\"o}k 2010, private communication). In this study, the extrapolated total magnetic field is regarded as a good approximation of the background magnetic fields. We do not believe that this approximation will affect the results of this study, which is largely based on the comparison between the relevant values for the two types of events.

However, an interesting finding, as shown in Figure \ref{index} is that, there appears an unusual ``bump" in the height distribution of the decay index \textit{n} around the height of $\sim$10 Mm (F7-F9) at the locations of eruptive events, whereas, for the confined flares, the ''bump" doesn't exist (F1--F6). At this height, the decay index is greater than 1.0 for eruptive flares F7 and F9, and approaches 1.0 for F8, whereas, for the confined flares, the decay index is apparently less than 1.0 for F3--F6 and also close to 1.0 for F1 and F2. In general, these results indicate that the transverse magnetic field in the lower corona over the eruptive flare sites decreases faster than that over the confined flare sites. Nevertheless, above the height of 40 Mm, the transverse magnetic field above the eruptive flares decreases slower than that for the confined flares.
Noted that, the decay index is almost the same at the height of $\sim$10 Mm for event F1, F2 and F8, however, the transverse field strength over the flare F8 site is weaker than that over the flare F1 and F2. More details for transverse field strength will be presented later. This difference decay properties of the background field may indicate that why the flares in the outskirt of the AR are eruptive; the eruption is helped by a larger decay index at the outskirts of the AR in the lower corona, where the flux rope is easier to experience the torus instability leading to the CME eruption. Recent theoretical analysis of \citet{olmedo10}, based on a more realistic assumption that the flux rope is a partial torus, show that the critical decay index is not a constant, but a function of the fraction of the torus about the surface. The critical index is smaller if the flux rope is lower lying, i.e., a flux rope in ARs.

Also noted that, the decay index profiles of NLFFF model is different from that of potential field model, especially in the low corona. Therefore, the potential field model may not be sufficient to characterize the decay index of the corona magnetic field. The NLFFF model based on observations of vector magnetic field is highly desirable for such studies, in particular for flare-productive ARs \citep[also see,][]{regnier02,Wie05,derosa09,schrijver09}.

\begin{figure*}  
     \vspace{0.0\textwidth}    
     \centerline{\hspace*{0.0\textwidth}
               \includegraphics[width=0.6\textwidth,clip=]{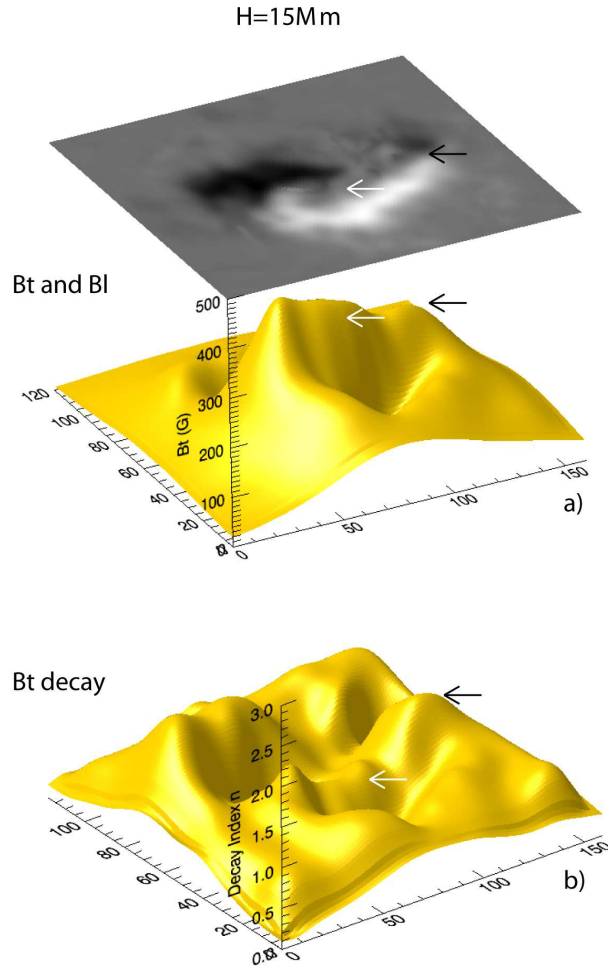}
              }
     \vspace{0.0\textwidth}   

\caption{Coronal transverse magnetic field at the height of 15 Mm of NOAA AR 10320 from extrapolated 3D magnetic field at 04:39 UT on 15 January. (a) The surface plot of the transverse magnetic field intensity (yellow surface plot). For the reference, the gray-scale image shows the line-of-sight magnetic field on the surface of the Sun. The white and black arrows denote the general locations of the confined and eruptive flares, respectively. Apparently, the outstanding ridge in the surface plot follows the PIL of the AR. (b) The surface plot of the decay index of the transverse magnetic field.}

   \label{iso}
   \end{figure*}

In order to further reveal the distinct magnetic properties between the eruptive and confined events at the low coronal height, we make the surface plot of the distribution of the transverse magnetic field over the AR at the height of 15 Mm (Figure \ref{iso}a). Apparently, the outstanding ridge follows the PIL of the AR. It shows that the strong transverse field mainly resides along the PIL. Departing from the PIL, the transverse field decreases. In the plot, the white arrow points at the general location of the confined flares, while the black arrow points at the location of the eruptive flares. The transverse field over the confined flares is evidently stronger than that above the eruptive flares. Further, we plot the decay index surface at the 15 Mm height (Figure \ref{iso}b). It is obvious that the decay index along the PIL is not a constant. The decay index is larger at the location of the eruptive flares (white arrow) than that of the confined flares (black arrow). One could argue that, the flux rope may exist in both the center and outskirts of the AR, however, it is more difficult to erupt if the flux rope situates at the center of the AR, because of the stronger overlying transverse magnetic field and the smaller decay index; the situation is opposite if the flux rope is located at the outskirt of the AR, it is easy to erupt due to the weaker transverse magnetic field and the larger decay index. Another possibility is that the flux rope extends from the center to the outskirt, but only the portion of the outskirt is erupted (partial eruption). This argument is consistent with our observations. This also explains why the confined flares have a smaller distance parameter $D$ than that of the eruptive ones.

\section{Discussions and Summary}
In addition to flux-rope-based CME models \citep{chen96,torok05,kliem06,fan07,olmedo10}, there are other CME models assuming a non-flux-rope magnetic structure prior to the eruption, including the tether-cutting model \citep{moore80,moore92}, the breakout model \citep{antiochos99}, the flux emergence model \citep{chen00}. In these models, the pre-eruption magnetic structure is a sheared core field instead of a twisted flux rope. Nevertheless, the sheared core field is believed to convert into a flux rope structure during the eruption through the magnetic reconnection mechanism. The highly sheared and elongated field lines along the PIL in our study, extrapolated by the NLFFF model, is suggestive of the existing of the flux rope over the AR. However, the flux rope seems only weakly twisted. We believe that the overlying magnetic field play an important role in determining its confinement and eruptiveness.


In flux rope models, different critical decay index values have been proposed. \citet{torok05} proposed a constant value around 1.5. \citet{olmedo10}, based on the analysis of the instability of a partial torus, suggested the critical index is a function of the fractional size of the flux rope; the critical value varies from 0 to 2 depending on the fraction of the torus above the surface. As shown in the previous section, the decay index inferred from our calculation in the low corona is large enough to cause the possible partial torus instability for eruptive flares, thanks to the bump-shaped distribution of the decay index with the height in the low corona.

Using a potential field model, \citet{guo102} recently studied the decay index distribution with the height for one confined flare, and found that the decay index is persistently smaller than 1.5 at the height range from $\sim$40 to 100 Mm. The filament started to rise at the height of $\sim$20 Mm, but as a consequence of the low decay index in the high corona, thus the absence of the torus instability, the erupting filament was confined and did not evolve into a CME. For the AR in our study, the decay index reaches a large value, i.e., $\sim$1.7, above the height of $\sim$60 Mm,  so that the torus instability will always occur in the high coronal. However, for the eruptive flares, the pre-eruption flux rope must undergo the instability in the low corona, possibly at the height of $\sim$10 Mm where the decay index is unusually large as inferred from the model calculation.  Subsequently, the erupting structure rose and reached a height that the torus instability will always occur, i.e., the index is larger than 1.5.

Besides the decay index, the transverse magnetic field strength also plays an important role in determining the confinement/eruptiveness of a flare.
Based on events from different ARs, \citet{liu08} found that the average transverse magnetic field strength for the confined events is about a factor of 3 stronger than that for the eruptive ones. In our study of the same AR, we find that the transverse magnetic field strength at the same height varies along the PIL; it is stronger near the AR center than the AR outskirts. The stronger transverse field provides the greater downward Lorentz force that keeps the flux rope from reaching a higher corona where the decay index is larger, thus makes the eruption mode difficult.

In conclusion, we have made a comparative study between the confined and eruptive flares that all occurred in the same NOAA AR 10720. We investigate their distinct properties in X-ray and EUV observations. Using the NLFFF model calculation based on vector magnetogram observations, we also investigate their distinct magnetic properties, in particular, the magnetic decay index. Our main findings are:

1, The confined flares tend to be more impulsive as seen in soft X-ray profiles, with a mean duration of 23 minutes, while the eruptive ones are mostly LDEs with a mean duration of 112 minutes.

2, The brightenings in the 195 \AA\ EIT images show distinct morphology. The confined flares display a more slender and compact shape, while the eruptive ones tend to be much more extensive in size.

3, The confined flares occur near the magnetic flux-weighted center of the AR, with a mean displacement of 16 Mm from the center, while the eruptive ones occur at the outskirts of the AR, with a mean displacement of 39 Mm. Nevertheless, all flares occurred near the PIL.

4, The decay index $n$ of the transverse magnetic field also have distinct properties for the two types of flares. For the eruptive events, a ``bump" appears at the height of $\sim$ 10 Mm in the height distributions of $n$, which indicates that the transverse field in the low corona decreases rapid enough to allow the eruption to occur. This is consistent with the theory of flux rope torus instability. Moreover, the transverse magnetic field strength for the eruptive events is weaker than that for the confined ones at the same heights.

In short, our study suggests that the farther the flare site from the magnetic flux center, the larger the possibility of the flare being an eruptive one. This is possibly related with the large magnetic decay index at these locations. Nevertheless, our study involves only a limited number events.  We plan to extend our study to a large number of events and make use of the advanced coronal and vector magnetic field observations of $Solar$ $Dynamic$ $Observatory$ ($SDO$).


\acknowledgements
We thank the anonymous referee for his/her valuable comments that helped to improve the paper. We are grateful to P. F. Chen, H. Li, and O. Olmedo for discussions and help in data analysis. X.C., M.D.D., and Y.G. were supported by NSFC under grants 10673004, 10828306, and 10933003 and NKBRSF under grant 2011CB811402. J.Z. was supported by NSF grant ATM-0748003 and NASA grant NNG05GG19G. SOHO is a project of international cooperation between ESA and NASA.





\begin{table*} 
\caption{Properties of flares in NOAA AR 10720.}
\label{tb1} \tabcolsep 0pt
\begin{tabular}{lcccccccccc}      
\\ \tableline \tableline                     

Flares~~&~Date~&~Onset~&~Rise~&~Duration~&~\textit{D}$^a$~&~Location~&~Class~&~CMEs$^b$~&~Speed$^c$~&~Width$^c$\\
~~~~~~~~&~---~&~(UT)~~&~(min)~&~(min)~&~(Mm)~&~~~---~~~~~&~~~---~~~~~&~~~---~~~&(km s$^{-1}$)&(deg)\\
\hline
\multicolumn{10}{c}{Confined Flares}\\
\hline
F1~~&~Jan 14~&~21:08~&~18~&~31~&~0.7~&~N14E10~&~M1.9~&~N~&~--~&~--\\
F2~~&~Jan 15~&~00:22~&~21~&~40~&~20.1~&~N14E08~&~X1.2~&~N~&~--~&~--\\
F3~~&~Jan 15~&~04:26~&~5~&~10~&~18.1~&~N14E06~&~M8.4~&~N~&~--~&~--\\
F4~~&~Jan 15~&~11:41~&~7~&~9~&~11.3~&~N14E02~&~M1.2~&~N~&~--~&~--\\
F5~~&~Jan 16~&~21:55~&~8~&~27~&~18.8~&~N15W19~&~M2.4~&~N~&~--~&~--\\
F6~~&~Jan 17~&~03:10~&~11~&~22~&~28.7~&~N15W21~&~M2.6~&~N~&~--~&~--\\
\hline
\multicolumn{10}{c}{Eruptive Flares}\\
\hline
F7~~&~Jan 15~&~05:54~&~44~&~83~&~31.9~&~N16E04~&~M8.6~&~Y~&~2049~&~360\\
F8~~&~Jan 15~&~22:25~&~37~&~66~&~34.5~&~N15W05~&~X2.6~&~Y~&~2861~&~360\\
F9$^d$~~&~Jan 17~&~06:59~&~173~&~188~&~49.7~&~N15W25~&~X3.8~&~Y (CME1)~&~2094~&~360\\
--~~~~&~~~--~~~&~~~--~~~&~~--~~&~~--~~&~~~--~~~&~~~--~~&~--~&~~~Y (CME2)~~~&~2547~&~360\\
\hline
\end{tabular}

\vspace{0.03\textwidth}
$^a$ \textit{D} denotes the distances between the intensity-weighted
flare center and the magnetic flux-weighted AR center.\\
$^b$ N denotes a confined flare that is not associated with a CME;
Y denotes an eruptive flare that is associated with a CME.\\
$^c$ The speed and angular width of CMEs are from http://cdaw.gsfc.nasa.gov/CME\_list.\\
$^d$ This extremely long duration flare is possibly related with two consecutive CMEs.\\
\end{table*}
\vspace{-0.0\textwidth}

\begin{table*} 
\caption{Flux Balance Parameters of the Boundary Condition In the NLFFF model}
\label{tb2} \tabcolsep 5pt
\begin{tabular}{lcccccc}     
\\ \tableline \tableline                    

Flares~~&~Date~&~Time$^a$~&~Interval$^b$~&~$\epsilon$$^c$~&~$\epsilon_{in}$$^d$~&~$\epsilon_{out}$$^e$~\\
~~~~~~~~&~----~&~(UT)~&~~~(Hr:Min)~~~&~~~~~---~~~~~~~&~~~---~~~~&~~---~~~~\\
\hline
\multicolumn{7}{c}{Confined Flares}\\
\hline
F1~~~~~&~Jan 14~&~~00:48~&~20:36~&~0.08~&~0.13~&~0.32~\\
F2~~~~~&~Jan 15~&~~00:39~&~00:09~&~0.17~&~0.07~&~0.54~\\
F3~~~~~&~Jan 15~&~~04:39~&~-00:03~&~0.18~&~0.06~&~0.53~\\
F4~~~~~&~Jan 15~&~~06:41~&~05:07~&~0.17~&~0.07~&~0.54~\\
F5~~~~~&~Jan 16~&~~03:58~&~18:02~&~0.03~&~0.16~&~0.24~\\
F6~~~~~&~Jan 17~&~~03:28~&~-00:04~&~--0.07~&~0.29~&~0.11~\\

\hline
\multicolumn{7}{c}{Eruptive Flares}\\
\hline
F7~~~~~&~Jan 15~&~~05:50~&~00:46~&~0.17~&~0.08~&~0.51~\\
F8~~~~~&~Jan 15~&~~06:41~&~16:31~&~0.17~&~0.07~&~0.54~\\
F9~~~~~&~Jan 17~&~~05:52~&~03:56~&~--0.06~&~0.15~&~0.12~\\

\hline
\end{tabular}
\vspace{0.03\textwidth}

$^a$ Observation time of SMFT vector magnetograms used in the NLFFF extrapolation.\\
$^b$ Time interval between the peak time of the flare and the
observational time of the vector magnetogram.\\
$^c$ The flux balance parameter at the bottom boundary
$\epsilon=\int_sB_zdxdy/\int_s\vert B_z \vert dxdy$, where s
denotes the bottom boundary.\\
$^d$ The inward flux ratio between the top and side boundaries and
the bottom boundary $\epsilon_{in}=(\int _{s1,s2} B_x^{in}
dydz+\int_{s3,s4} B_y^{in} dxdz+\int_{s5} B_z^{in} dxdy)/ \int_s
B_z^{in} dxdy$, where s1--s5 denote the five boundaries of the
potential field that is extrapolated using the Green function method
based on the bottom boundary.\\
$^e$ The outward flux ratio between the top and side boundaries and
the bottom boundary $\epsilon_{out}=(\int _{s1,s2} B_x^{out}
dydz+\int_{s3,s4} B_y^{out} dxdz+\int_{s5} B_z^{out} dxdy)/ \int_s
B_z^{out} dxdy$.

\end{table*}


\end{document}